\documentclass[conference]{IEEEtran}
\IEEEoverridecommandlockouts

\usepackage{amsmath,amssymb}
\usepackage{booktabs}
\usepackage{graphicx}
\usepackage{multirow}
\usepackage{url}
\usepackage[normalem]{ulem}

\begin{document}

\title{How Far Can Wearable-Compatible Signals Go? A Controlled Decomposition of Non-EEG Sleep Staging}

\author{\IEEEauthorblockN{Yi Wang}%
\thanks{This work has been submitted to the IEEE for possible publication. Copyright may be transferred without notice, after which this version may no longer be accessible.}}

\maketitle

\begin{abstract}

Consumer wearables increasingly infer sleep stages from signals including heart rate, accelerometry, and photoplethysmography. However, existing studies often report end-to-end performance under a fixed signal setting, making it difficult to determine whether the observed performance comes from genuine physiological decoding, temporal priors, or dataset-specific confounds. To address this limitation, we introduce a four-layer controlled decomposition framework for non-EEG sleep staging, covering signal source, physiological representation, temporal prior, and decision layers. The framework is evaluated across a signal-quality ladder spanning Apple Watch Sleep-Accel ($N=31$), the Sleep Heart Health Study ($N=195$, laboratory ECG, respiratory, and SpO$_2$ signals), and Sleep-EDF-20 as an EEG+EOG reference, using the same compact Mamba2 model~\cite{gu2023mamba,dao2024transformers} throughout. Laboratory cardiorespiratory signals reach $\kappa=0.492$, while EEG+EOG reaches $\kappa=0.796$, leaving a residual gap of $\Delta\kappa=+0.304$ that reflects missing cortical information rather than temporal modeling alone. Consumer HR/ACC reaches only $\kappa=0.255$, quantifying the additional penalty of derived wearable signals and real-world sensing constraints. Confidence-based abstention provides a calibrated operating mode: removing the 20\% lowest-confidence epochs increases $\kappa$ from $0.452$ to $0.512$, while a label-shuffled control collapses to $\kappa=-0.003$. These results support non-EEG sleep staging as coarse, confidence-aware sleep-structure monitoring rather than EEG-equivalent five-class clinical staging.

\end{abstract}


\section{Introduction}

Clinical sleep staging relies on electroencephalography (EEG)~\cite{zhang2025harnessing,yu2025machine}, electrooculography (EOG)~\cite{zibandehpoor2025electrooculography}, and electromyography (EMG)~\cite{boo2025comprehensive,liu2026pose} recorded during attended polysomnography (PSG)~\cite{leger2025polysomnography}. Trained technologists annotate 30-second epochs into Wake, N1, N2, N3, and REM according to AASM criteria~\cite{nieto2026severity}, yielding inter-rater agreement typically in the range $\kappa \approx 0.80$--$0.85$ when scored against a consensus reference~\cite{berry2017aasm}. Automated EEG-based systems now approach similar levels, with reported $\kappa$ values frequently exceeding $0.75$ on benchmark datasets such as Sleep-EDF and SHHS~\cite{supratak2017deepsleepnet,perslev2021u,phan2022xsleepnet}.

Consumer wearable devices, including wrist-worn actigraphs, smartwatches, and finger-worn rings, estimate sleep stages from a fundamentally different signal set~\cite{doherty2025privacy}. Instead of directly observing cortical electrophysiology, these devices typically rely on photoplethysmography (PPG)-derived heart rate and heart-rate variability, triaxial accelerometry, and, in some configurations, peripheral oxygen saturation or respiration-related surrogates derived from PPG modulation. These signals capture downstream autonomic and respiratory correlates of sleep stage: sympathetic--parasympathetic balance changes across NREM and REM sleep, cardiac inter-beat intervals reflect respiratory sinus arrhythmia, and body movement co-varies with wakefulness and stage transitions~\cite{tobaldini2013heart,trinder2001autonomic}. The central question is therefore not whether non-EEG correlates of sleep exist, but how much clinically meaningful sleep-stage structure can be recovered when cortical signals are absent.

Prior work has demonstrated above-chance non-EEG sleep staging using heart-rate variability, respiratory signals, SpO$_2$, and actigraphy features, with reported Cohen's $\kappa$ values typically ranging from approximately $0.30$ to $0.55$, depending on signal availability, cohort characteristics, and evaluation protocol~\cite{sridhar2020deep,radha2019sleep,sun2019cardioresp}. However, most studies report a single aggregate metric from a fixed end-to-end pipeline. Such reporting makes it difficult to determine whether performance arises from genuine physiological discriminability, improved feature representation, temporal smoothing, model capacity, or evaluation artifacts. For example, a reported $\kappa=0.45$ could reflect useful autonomic and respiratory information, an overly strong temporal prior, subject-level leakage, or a combination of these factors. A more diagnostic evaluation requires separating the sleep-staging pipeline into independently testable components.

This study introduces a controlled decomposition framework for non-EEG sleep staging. The framework separates the pipeline into four layers: signal source, physiological representation, temporal prior, and decision. The signal-source layer evaluates what information is available under different sensing conditions. The representation layer evaluates how raw or derived signals are transformed into sleep-relevant physiological features. The temporal-prior layer measures how much sequence decoding contributes beyond per-epoch discriminability. The decision layer evaluates whether the model should report a sleep stage or abstain when confidence is low. By holding the model architecture and evaluation protocol fixed while varying these layers, the framework attributes performance gains and residual failures to specific parts of the pipeline rather than reporting a single undifferentiated score.

We evaluate the decomposition across a signal-quality ladder spanning consumer wearable signals, laboratory PSG-derived non-EEG physiology, and EEG+EOG reference staging. The consumer tier uses Apple Watch heart rate and accelerometry. The laboratory non-EEG tier uses ECG, respiratory, and SpO$_2$ signals from SHHS. The EEG+EOG tier uses Sleep-EDF-20 as a reference ceiling for direct cortical and ocular measurements. Across tiers, we use an identical compact state-space model with multi-scale temporal evidence aggregation and a fixed evaluation protocol. We further include subject-disjoint validation, label-shuffled negative controls, subject-overlap audits, and bootstrap confidence intervals.

Rather than proposing another state-of-the-art sleep-staging model, this work contributes a methodology for diagnosing the performance limits of non-EEG sleep staging. The main contributions are as follows:
\begin{enumerate}
\item A four-layer controlled decomposition framework is introduced, separating non-EEG sleep staging into signal source, physiological representation, temporal prior, and decision layers, with the marginal contribution of each layer quantified independently.
\item Physiological representation provides the largest recoverable gain within the non-EEG pipeline ($\Delta\kappa=+0.078$), while temporal decoding contributes modestly ($\Delta\kappa=+0.040$), indicating that the binding constraint for wearable-compatible non-EEG staging is the per-epoch physiological information content of the signals themselves, not the capacity of the temporal model---deeper sequence models cannot compensate for limited signal physiology.
\item Confidence-based abstention separates reliable epochs from uncertain ones: at $50\%$ coverage the model reaches $\kappa=0.62$, and the per-class confidence pattern faithfully reflects the physiological ambiguity of transitional N1 sleep.
\item The residual gap to EEG+EOG performance ($\Delta\kappa=+0.304$) is attributable to signal modality rather than model capacity, and exceeds the combined gain of all non-EEG layers.
\end{enumerate}

\section{Related Work}

\subsection{EEG-Based Automatic Sleep Staging}

Most high-performing automatic sleep-staging systems use EEG, EOG, EMG, or combinations of PSG channels. DeepSleepNet uses raw single-channel EEG with representation learning and sequence modeling~\cite{supratak2017deepsleepnet}. TinySleepNet studies efficient EEG staging~\cite{supratak2020tinysleepnet}. U-Sleep emphasizes cross-dataset robustness and high-frequency staging from PSG-like signals~\cite{perslev2021u}. XSleepNet and SleepTransformer further model sequential and multi-view sleep structure~\cite{phan2022xsleepnet,phan2022sleeptransformer}. These systems define the EEG-rich reference ceiling, reflecting direct cortical sleep physiology. Cardiac and respiratory signals reflect downstream autonomic and respiratory correlates, operating on fundamentally different information sources. Comparing the two domains without documenting signal access conflates information content with model capacity.

\subsection{Wearable and Non-EEG Sleep Staging}

Wearable sleep staging typically relies on motion, heart rate, HRV, PPG/BVP, respiration, or SpO$_2$~\cite{imtiaz2021systematic}. Walch et al. studied Apple Watch acceleration and PPG-derived heart rate against PSG and reported strong sleep/wake performance but limited stage resolution~\cite{Walch2019SleepAccel}. Broader consumer-device validation work reaches a similar conclusion: sleep detection sensitivity is often high, but wake detection and stage assessment are more variable~\cite{chinoy2021seven,altini2021promise}.

Non-EEG staging performance improves when the signal source is closer to laboratory physiology. Heart-rate-based staging with ECG-derived instantaneous heart rate has been demonstrated on large clinical datasets~\cite{sridhar2020deep,radha2019sleep}. ECG combined with respiratory effort can reach substantially stronger agreement in PSG-derived cohorts~\cite{sun2019cardioresp,willemen2014evaluation}. Raw PPG time-series models represent a different signal condition from the sparse, device-derived HR values commonly exported by consumer watches~\cite{kotzen2022sleepppg}. Critically, these studies rarely report the marginal contribution of each pipeline component, making it impossible to distinguish strong signal decoding from strong temporal priors or data leakage.

\subsection{Controlled Decomposition Methodology}

The principle of controlled decomposition---separating a physiological inference problem into its constituent layers and evaluating each independently---has been productive in robust wearable sensing. In camera-based and PPG-based heart rate estimation under motion, decomposing the problem into candidate generation, consensus selection, and temporal tracking has yielded interpretable performance boundaries and actionable diagnoses of failure modes. We adapt this principle to non-EEG sleep staging by partitioning the pipeline into four independently evaluated layers: signal source, physiological representation, temporal prior, and decision. Each layer's marginal contribution is quantified as the $\Delta\kappa$ obtained when that layer is added to the previous layers, and residual failure is attributed to the layer where improvement saturates. This decomposition is especially important in sleep staging because adjacent epochs are temporally autocorrelated, stage labels are strongly imbalanced, and subject identity can leak through sensor-specific artifacts. A model that exploits dataset-specific transition statistics, subject-identity leakage, or annotation shortcuts can report an attractive aggregate metric without capturing physiological sleep-stage information. A defensible wearable-facing evaluation therefore requires channel ablations, subject-disjoint splits, temporal decoding comparisons, and negative controls---each isolating one layer of the decomposition.

\section{Methods}

\subsection{Framework}

\begin{figure*}
    \centering
    \includegraphics[width=1\linewidth]{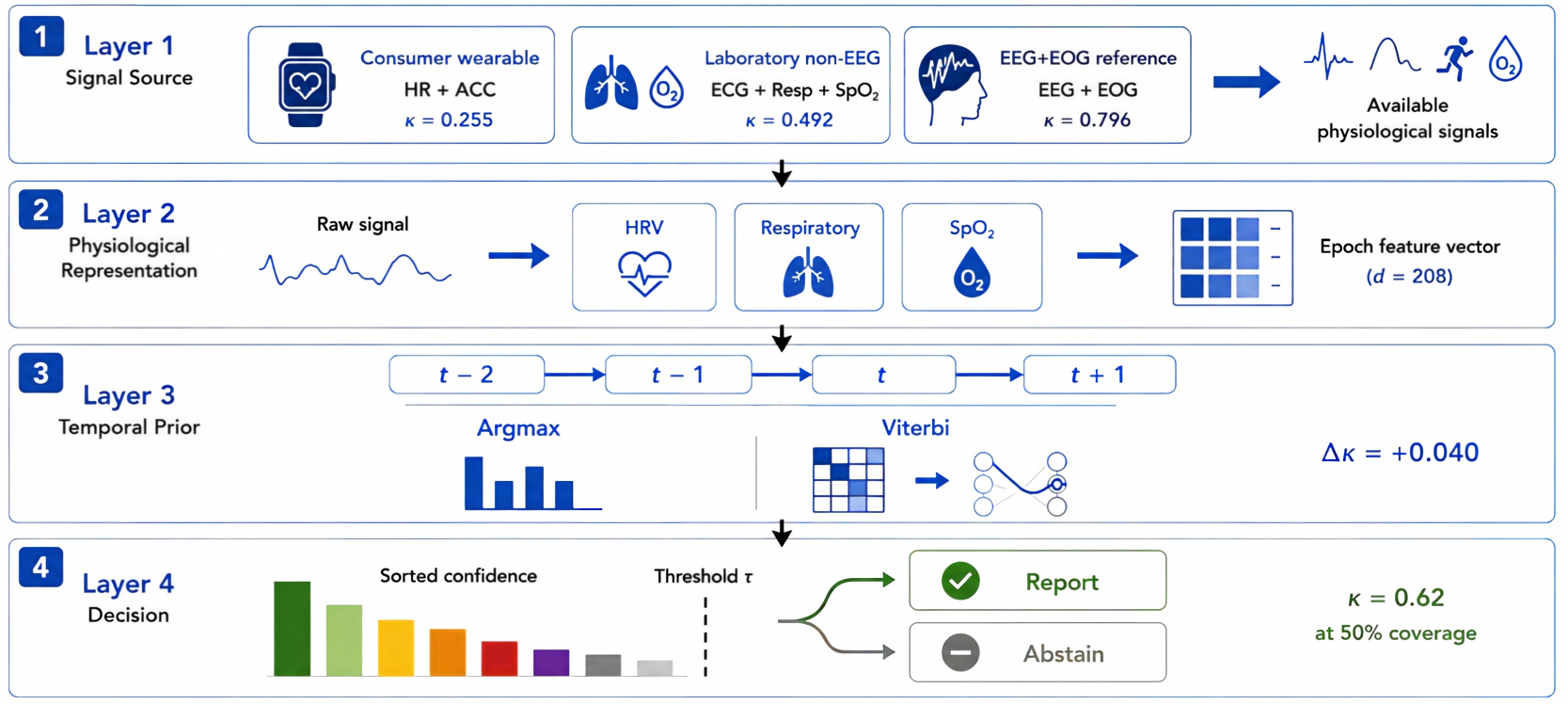}
\caption{Overview of the controlled decomposition framework for non-EEG sleep staging. The framework separates the staging pipeline into four independently evaluated layers: signal source, physiological representation, temporal prior, and decision. Signals are evaluated across a quality ladder from consumer wearable HR/ACC, to laboratory ECG, respiratory, and SpO$_2$ measurements, and finally to EEG+EOG reference staging. By fixing the model architecture and evaluation protocol across tiers, the framework attributes performance differences to physiological information, temporal decoding, confidence-based abstention, and residual signal-modality limitations.}
    \label{fig:framework}
\end{figure*}

Figure~\ref{fig:framework} summarizes the proposed controlled decomposition framework for non-EEG sleep staging. Instead of evaluating a non-EEG staging system as a single end-to-end classifier, the framework separates performance into four experimentally controlled layers: signal source, physiological representation, temporal prior, and decision.

The signal-source layer measures the information available from different sensing conditions. We compare consumer wearable signals, laboratory-grade non-EEG PSG signals, and EEG+EOG reference signals while keeping the downstream model and evaluation protocol fixed. The physiological-representation layer evaluates how much stage-relevant information can be extracted from the available non-EEG signals, including cardiac rhythm, HRV, respiratory dynamics, thoracoabdominal coupling, and oxygen desaturation features. The temporal-prior layer quantifies the contribution of sleep-stage transition structure by comparing independent per-epoch argmax decoding with Viterbi sequence decoding. The decision layer evaluates whether the model can identify reliable and unreliable predictions through confidence-based abstention.

This design allows each performance gain to be attributed to a specific source. Improvements from richer non-EEG features reflect recoverable physiological information; improvements from Viterbi decoding reflect temporal regularization; and the remaining gap to EEG+EOG reference performance reflects information unavailable to wearable-compatible non-EEG signals.

\subsection{Datasets and Labels}

Three datasets spanning a signal-quality ladder were used.

\textbf{Apple Watch Sleep-Accel}~\cite{Walch2019SleepAccel} served as the consumer-device tier. The public release contains Apple Watch heart rate (HR), triaxial accelerometry (ACC), and PSG-derived sleep-stage labels for 31 subjects, totaling $26{,}417$ aligned 30-second epochs. Labels follow standard five-class AASM staging (Wake, N1, N2, N3, REM). Evaluation was leave-one-subject-out (LOSO).

\textbf{Sleep Heart Health Study (SHHS)}~\cite{quan1997sleep,zhang2018nsrr} served as the laboratory non-EEG tier. SHHS is a multi-center community cohort with attended PSG, including ECG, respiratory inductance plethysmography (thoracic and abdominal), nasal airflow, and finger pulse oximetry. From the SHHS Visit 1 records, $200$ subjects with complete ECG R-peak annotations were identified. Of these, $195$ also had complete respiratory and SpO$_2$ recordings, forming the matched cohort used for channel-level ablation (denoted rpoint200). Five-fold subject-disjoint cross-validation was used.

\textbf{Sleep-EDF-20}~\cite{kemp2000analysis,goldberger2000physiobank} served as the EEG+EOG reference tier. This public dataset contains $20$ healthy subjects with frontal EEG (Fpz-Cz) and EOG (horizontal). Evaluation was leave-one-subject-out. This tier is used exclusively as a cross-dataset upper anchor; it is not a same-dataset baseline.

\subsection{Signal Representations and the Four-Layer Pipeline}

\textbf{Layer 1---Signal Source.} Three input channel configurations were evaluated on the SHHS matched cohort: ECG only ($d=133$), Respiration/SpO$_2$ only ($d=75$), and Combined ($d=208$). For the Apple Watch tier, HR-only, ACC-only, and HR+ACC configurations were evaluated.

\textbf{Layer 2---Physiological Representation.} All SHHS features were physiology-aware by construction. ECG rhythm features included instantaneous heart rate (IHR) and inter-beat interval (IBI) sequences interpolated to a $2$ Hz grid (120 values), plus 13 epoch-level HRV statistics (SDNN, RMSSD, pNN50, etc.)~\cite{pan1985real}. Respiratory and SpO$_2$ event features included per-channel respiratory rate (zero-crossing and peak-detection), amplitude statistics, flat/low-amplitude fractions, slope irregularity, and thoracic-abdominal coupling features (correlation, paradoxical-breathing fraction, amplitude ratio). SpO$_2$ features included eight per-epoch desaturation statistics. All features were robustly normalized per subject (median and MAD) with statistics computed on training folds only. Apple Watch features used epoch-level HR and ACC statistics as exported by the consumer device, representing the information available under real-world constraints without access to raw sensor waveforms.

\textbf{Layer 3---Temporal Prior.} Per-epoch argmax classification was supplemented with Viterbi sequence decoding~\cite{viterbi1967error}. Transition log-probabilities were estimated from training-fold labels with Laplace smoothing ($\alpha=1.0$). The Viterbi path was computed as:
\begin{equation}
\hat{\mathbf{y}}_{1:T} = \arg\max_{\mathbf{y}_{1:T}} \sum_{t=1}^{T} \left[ \log P(y_t \mid \mathbf{x}_t) + \lambda \cdot \log P(y_t \mid y_{t-1}) \right]
\end{equation}
with $\lambda \in \{0.1, 0.3, 0.5, 1.0\}$. The argmax result ($\lambda=0$) isolates per-epoch physiological discriminability. The best Viterbi result per channel configuration is reported. Transition statistics were estimated exclusively from training folds.

\textbf{Layer 4---Decision.} Per-epoch softmax probabilities were retained from the trained model. Epochs were ranked by maximum class probability, and Cohen's $\kappa$ was computed at coverage levels from 10\% to 100\%, where coverage denotes the fraction of highest-confidence epochs retained. This coverage-$\kappa$ curve characterizes the trade-off between prediction coverage and staging agreement: the system can abstain on low-confidence epochs to achieve higher agreement on the retained subset.

\subsection{Model Architecture}

A compact state-space sequence model (SSM) was used identically across all signal tiers. The architecture consisted of a linear input projection ($d_{\text{in}} \rightarrow d_{\text{model}}=64$), two Mamba2 blocks with state dimension $d_{\text{state}}=16$ and head dimension $d_{\text{head}}=8$, followed by layer normalization and a multi-scale temporal evidence aggregation (Multi-MEA) readout head. The Multi-MEA head applied four parallel 1D depthwise convolutions with kernel sizes $k \in \{1, 3, 5, 7\}$ and independent gating (sigmoid), each followed by a linear classifier. The four outputs were averaged to produce the final per-epoch logits.

The model was trained with AdamW (learning rate $3 \times 10^{-3}$, weight decay $10^{-4}$), inverse-frequency class weights, and cross-entropy loss. Training ran for $10$ epochs per fold with gradient clipping (max norm $1.0$). All experiments used seed $42$. Total parameter count was approximately $1.2 \times 10^5$, identical across tiers.

\begin{table}[t]
\centering
\footnotesize
\caption{Feature-space separability of five sleep stages in the 201-dimensional combined non-EEG feature space ($N=196{,}887$ epochs, $195$ subjects). ``Margin'' is the Euclidean distance to the nearest other stage's centroid. ``Ratio'' is margin divided by mean within-stage dispersion. LDA AUC is the one-vs-rest area under the ROC curve.}
\label{tab:separability}

\begin{tabular}{@{}lcccc@{}}
\toprule
Stage & Margin & Within $\sigma$ & Ratio & LDA AUC \\
\midrule
Wake & 10.6 & 18.9 & 0.558 & 0.896 \\
N1    & \bfseries 1.2 & 9.2  & \bfseries 0.133 & \bfseries 0.684 \\
N2    & 1.2  & 7.6  & 0.161 & 0.734 \\
N3    & 1.6  & 7.2  & 0.228 & 0.745 \\
REM   & 1.6  & 8.2  & 0.200 & 0.753 \\
\bottomrule
\end{tabular}%

\end{table}

\subsection{Evaluation Protocol and Verification}

The primary metric was Cohen's $\kappa$~\cite{cohen1960coefficient}, chosen over accuracy because it corrects for chance agreement from imbalanced stage distributions. Secondary metrics included macro-averaged $F_1$, per-stage $F_1$, and confusion matrices. Per-subject $\kappa$ values were computed with 95\% bootstrap confidence intervals via $10{,}000$ iterations of subject-level resampling.

A label-shuffled negative control was performed: sleep-stage labels were randomly permuted within each subject's record before the fold split (seed $42$), and the identical training protocol was executed. Under the null hypothesis, $\kappa$ should be indistinguishable from zero. A subject-overlap audit verified zero cross-fold subject contamination across all SHHS experiments.

\subsection{Experimental Workflow}

\begin{figure*}
    \centering
    \includegraphics[width=1\linewidth]{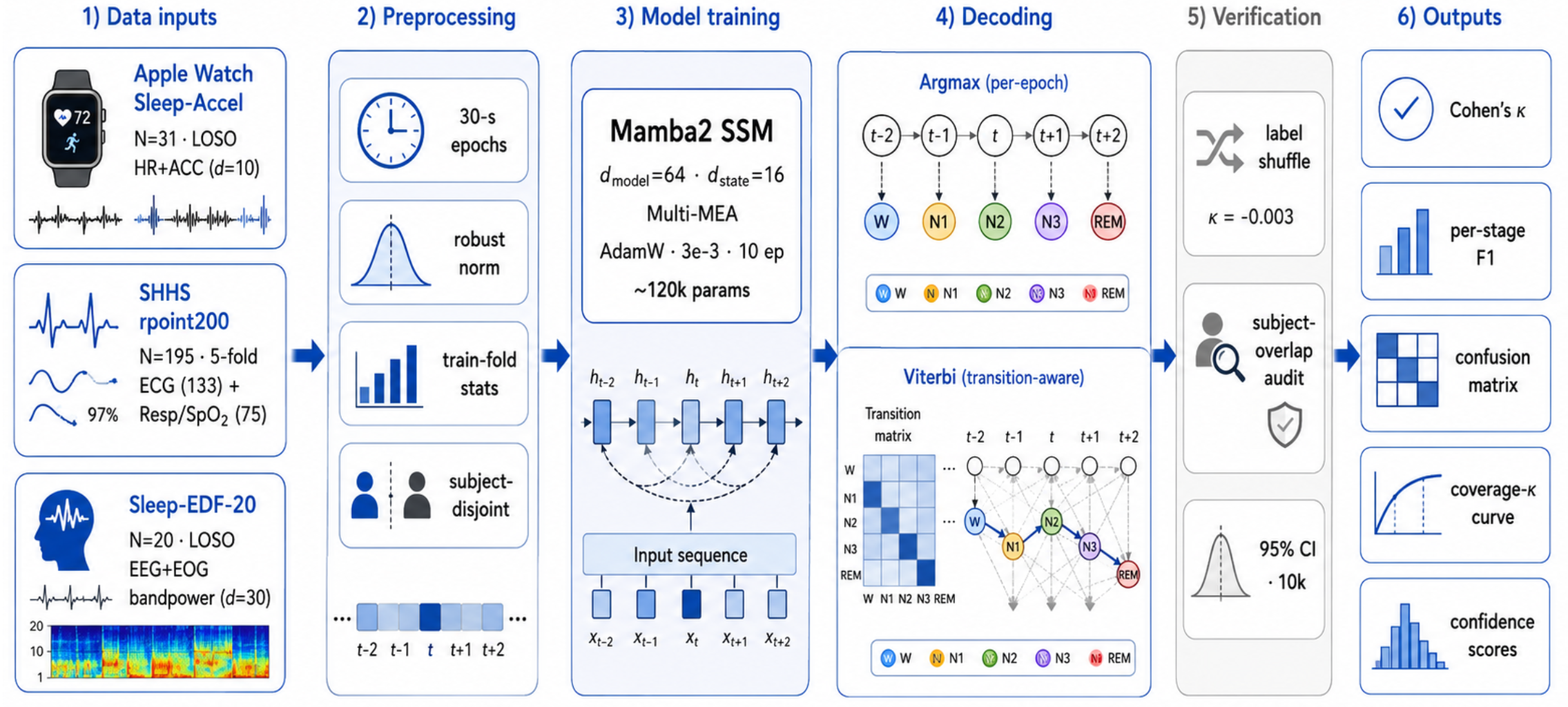}
    \caption{Experimental workflow for the controlled decomposition study. Signals from consumer wearable, laboratory non-EEG, and EEG+EOG reference tiers are aligned to 30-second sleep-stage epochs, transformed into physiological feature representations, and evaluated under subject-disjoint validation. The same compact state-space model is trained across all signal configurations. Predictions are decoded using argmax and Viterbi inference, followed by metric computation, confidence-based abstention analysis, and verification through label-shuffled negative controls, subject-overlap audits, and bootstrap confidence intervals.}
    \label{fig:workflow}
\end{figure*}

Figure~\ref{fig:workflow} illustrates the experimental workflow. For each dataset tier, signals were aligned to 30-second sleep-stage epochs and converted into the corresponding feature representation. Subject-disjoint splits were then constructed to prevent subject-level leakage. Within each training fold, normalization statistics, class weights, and transition probabilities were estimated using training subjects only. The same compact state-space model was trained across all signal configurations. Test predictions were evaluated using both argmax decoding and Viterbi decoding, followed by Cohen's $\kappa$, macro-$F_1$, per-stage $F_1$, confusion-matrix analysis, and confidence-based abstention. Label-shuffled controls, subject-overlap audits, and bootstrap confidence intervals were used to verify that the observed performance reflected physiological information rather than leakage or chance agreement.

\section{Results}

\subsection{Four-Layer Decomposition}

\begin{table*}[t]
\centering
\footnotesize
\caption{Four-layer decomposition of non-EEG sleep staging on SHHS ($N=195$, five-fold subject-disjoint). Each row adds one layer to the pipeline. $\Delta\kappa$ is the marginal gain over the cumulative result of all previous layers.}
\label{tab:decomposition}
\begin{tabular}{@{}lccc@{}}
\toprule
Layer added & Cumulative $\kappa$ & $\Delta\kappa$ & Interpretation \\
\midrule
1. ECG physiology alone (Viterbi) & 0.403 & --- & Baseline: cardiac autonomic info \\
2. + Resp/SpO$_2$ physiology & 0.452 & +0.049 & Respiratory \& oximetry info added \\
3. + Viterbi temporal prior & 0.492 & +0.040 & Modest temporal smoothing gain \\
4. EEG/EOG ceiling (Sleep-EDF-20) & 0.796 & +0.304 & Irrecoverable: cortical EEG missing \\
\bottomrule
\end{tabular}
\end{table*}

Table~\ref{tab:decomposition} presents the core result: the marginal contribution of each pipeline layer to non-EEG sleep staging performance on the SHHS matched cohort. Three patterns are evident. First, physiological representation provides the largest gain: expanding from cardiac autonomic features alone to combined ECG, respiratory, and SpO$_2$ features yields $\Delta\kappa=+0.078$ in argmax (ECG argmax $\kappa=0.373$ $\rightarrow$ Combined argmax $\kappa=0.452$), or $\Delta\kappa=+0.089$ in Viterbi. This confirms that cardiac and respiratory physiology carry complementary but non-redundant sleep-stage information. Second, the temporal prior contributes only $\Delta\kappa=+0.040$, a small gain relative to the representation gain. This is diagnostic: it indicates that the bottleneck is not the absence of temporal context but the limited sleep-stage information in autonomic and respiratory surrogates per epoch. Third, the residual gap to EEG+EOG staging ($\Delta\kappa=+0.304$) is larger than the entire range spanned by all non-EEG layers combined ($+0.089$), demonstrating that the dominant performance ceiling is signal modality, not model capacity or temporal modeling.

\subsection{Channel Ablation}

Table~\ref{tab:shhs_channel} reports per-channel performance. ECG rhythm alone and Resp/SpO$_2$ alone yield comparable five-class $\kappa$ ($0.403$ and $0.419$ in Viterbi, respectively), indicating that neither modality dominates non-EEG sleep-stage information. The combined representation outperforms either alone, confirming complementarity. N1 $F_1$ remains below $0.20$ for all channel configurations, indicating that N1's poor discriminability is not channel-specific but inherent to its weak autonomic signature.

\begin{table}[t]
\centering
\footnotesize
\caption{SHHS channel ablation ($N=195$, five-fold subject-disjoint). Per-subject argmax $\kappa$ with 95\% bootstrap CI. Viterbi at $\lambda=1.0$.}
\label{tab:shhs_channel}
\resizebox{\columnwidth}{!}{%
\begin{tabular}{@{}lccccc@{}}
\toprule
Channel set & $d$ & Argmax $\kappa$ & Per-subj.\ $\kappa$ [95\% CI] & Viterbi $\kappa$ & N1 $F_1$ \\
\midrule
ECG + IHR/IBI          & 133 & 0.373 & 0.369 $[0.353, 0.384]$ & 0.403 & 0.185 \\
Resp/SpO$_2$ + events  & 75  & 0.388 & 0.381 $[0.367, 0.396]$ & 0.419 & 0.155 \\
Combined               & 208 & 0.452 & 0.445 $[0.430, 0.460]$ & \textbf{0.492} & 0.20 \\
\bottomrule
\end{tabular}%
}
\end{table}

\subsection{Abstention Analysis: Confidence-Calibrated Staging}

\begin{figure}[t]
\centering
\includegraphics[width=\columnwidth]{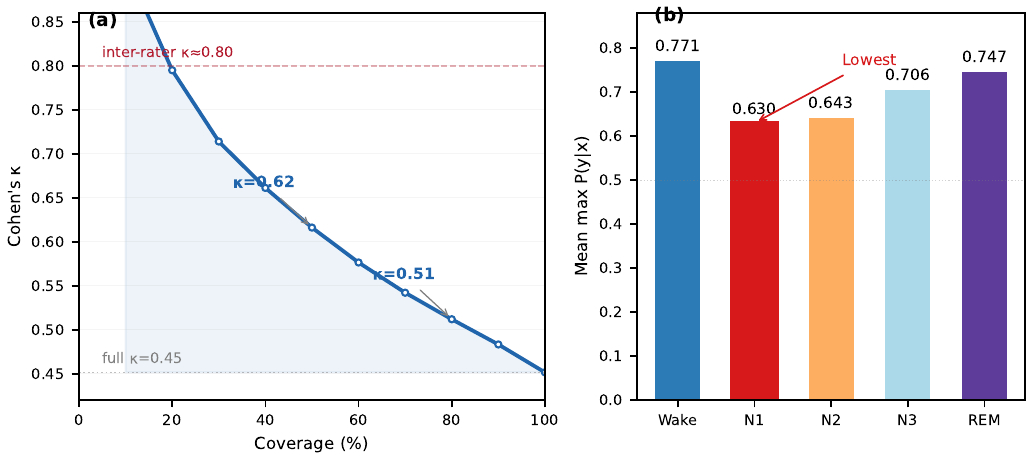}
\caption{Coverage-$\kappa$ curve for the SHHS combined model (argmax). Epochs are sorted by decreasing softmax confidence. Shaded region: 95\% bootstrap CI. Dashed horizontal lines: EEG+EOG reference ($\kappa=0.796$) and technologist inter-rater agreement ($\kappa\approx0.80$).}
\label{fig:abstention}
\end{figure}

Figure~\ref{fig:abstention} shows the coverage-$\kappa$ curve: epochs are ranked by model confidence (maximum softmax probability), and $\kappa$ is computed at decreasing coverage levels. The model is well-calibrated: discarding the 20\% lowest-confidence epochs raises $\kappa$ from $0.452$ to $0.512$; at 50\% coverage, $\kappa$ reaches $0.616$; at 30\% coverage, $\kappa=0.714$. The abstention gain is monotonic and substantial, confirming that the model's per-epoch confidence is informative rather than arbitrary.

Per-class confidence analysis reveals that the model's uncertainty is physiologically structured, not random. Table~\ref{tab:confidence} reports mean maximum softmax probability per sleep stage. Wake receives the highest mean confidence ($0.771$) and N1 the lowest ($0.630$). This ordering---Wake $>$ REM $>$ N3 $>$ N2 $>$ N1---matches the known strength of autonomic differentiation across sleep stages: Wake is characterized by elevated sympathetic tone and irregular respiration, producing a strong and stable non-EEG signature; N1 is transitional, short-duration, and poorly differentiated from relaxed Wake or light N2 in both EEG and autonomic physiology. The fact that model confidence tracks this physiological hierarchy, rather than simply reflecting class prevalence (N2 is the majority class yet receives only moderate confidence, $0.643$), indicates that the abstention signal is genuine rather than a prevalence artifact.

\begin{table}[t]
\centering
\caption{Per-stage mean model confidence (max softmax probability).}
\label{tab:confidence}
\begin{tabular}{lcc}
\toprule
Stage & Mean confidence & $N$ epochs \\
\midrule
Wake & 0.771 & 50{,}626 \\
N1    & 0.630 & 7{,}349 \\
N2    & 0.643 & 82{,}853 \\
N3    & 0.706 & 25{,}948 \\
REM   & 0.747 & 30{,}111 \\
\bottomrule
\end{tabular}
\end{table}

N1 is the clearest failure mode: 21.9\% of N1 epochs receive a maximum class probability below $0.5$, compared to $7.0\%$ of Wake epochs. When epochs with confidence below $0.7$ are excluded, overall $\kappa$ rises to $0.622$ but only $33.6\%$ of N1 epochs survive the filter. This is precisely the expected behavior under a physiological bottleneck: the model cannot confidently classify N1 because the autonomic and respiratory signatures of N1 are intrinsically weak, and it appropriately assigns low confidence to those epochs. The correct operational conclusion is not that the model should be improved to better classify N1, but that N1 should be excluded from non-EEG staging reports or merged into a Light Sleep category, and that low-confidence epochs of any stage should be flagged for human review or discarded.

\subsection{Per-Stage Failure Analysis}

\begin{figure}[t]
\centering
\includegraphics[width=0.7\columnwidth]{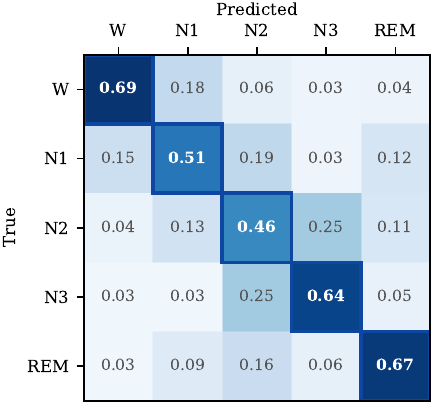}
\caption{Normalized confusion matrix for SHHS combined model (pooled five-fold, argmax). Rows are true stages, columns are predicted stages. Each row sums to 1.0. Off-diagonal entries $>0.15$ are annotated.}
\label{fig:confusion}
\end{figure}

Table~\ref{tab:per_stage} reports per-stage $F_1$ for the SHHS combined Viterbi model. Four-class aggregation (Wake, Light, Deep, REM) raised Viterbi $\kappa$ to $0.517$, confirming that the dominant source of confusion is the N1/N2 boundary. A normalized confusion matrix (Figure~\ref{fig:confusion}) shows that N1 is systematically absorbed into N2 ($19\%$) and Wake ($15\%$), quantitatively confirming the transitional nature of N1 under non-EEG physiology.

\begin{table}[t]
\centering
\caption{Per-stage $F_1$ for SHHS combined Viterbi model ($\lambda=1.0$, $N=195$).}
\label{tab:per_stage}
\begin{tabular}{lccccc}
\toprule
& W & N1 & N2 & N3 & REM \\
\midrule
$F_1$ & 0.78 & 0.20 & 0.61 & 0.52 & 0.66 \\
\bottomrule
\end{tabular}
\end{table}

\subsection{Consumer-Device Context}

\begin{figure}[t]
\centering
\includegraphics[width=\columnwidth]{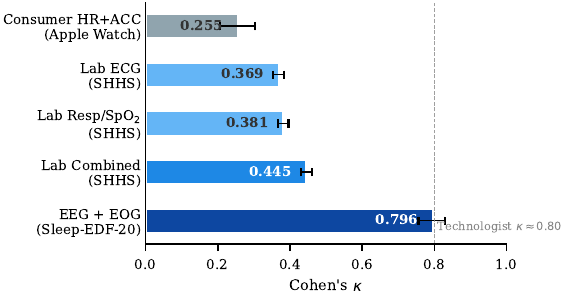}
\caption{Signal-availability ladder. Horizontal bars show per-subject Cohen's $\kappa$ with 95\% bootstrap confidence intervals. Dashed lines mark technologist inter-rater agreement ($\kappa\approx0.80$) and the empirical non-EEG ceiling ($\kappa\approx0.49$).}
\label{fig:signal_ladder}
\end{figure}

The signal-quality ladder (Figure~\ref{fig:signal_ladder}) places the laboratory cardiorespiratory results in context with consumer-device and EEG reference performance. The Apple Watch HR+ACC tier achieves $\kappa=0.255$, approximately half the laboratory combined result, consistent with the reduced physiological information in consumer-exported HR (5-second sampling) and the limited sleep-stage information in wrist accelerometry (Table~\ref{tab:apple_watch}). The gap from consumer HR+ACC to laboratory combined non-EEG ($\Delta\kappa=0.24$) and the gap from combined non-EEG to EEG+EOG ($\Delta\kappa=0.30$) are of comparable magnitude, indicating that both signal quality and signal modality contribute substantially to the overall PSG-to-wearable performance gap.

\begin{table}[t]
\centering
\footnotesize
\caption{Apple Watch Sleep-Accel consumer-device results (31-subject LOSO).}
\label{tab:apple_watch}
\begin{tabular}{@{}lcccc@{}}
\toprule
Input & 5-cl.\ $\kappa$ [95\% CI] & 5-cl.\ MF1 & 3-cl.\ $\kappa$ & 3-cl.\ MF1 \\
\midrule
ACC-only & 0.154 $[0.124, 0.185]$ & 0.338 & 0.159 & 0.483 \\
HR-only  & 0.270 $[0.223, 0.314]$ & 0.414 & 0.349 & 0.566 \\
HR+ACC   & 0.255 $[0.206, 0.304]$ & 0.414 & 0.332 & 0.565 \\
\bottomrule
\end{tabular}
\end{table}

\subsection{Negative Control}

The label-shuffled control on the SHHS combined experiment produced five-class argmax $\kappa=-0.003$, indistinguishable from chance. Viterbi decoding at $\lambda=1.0$ yielded $\kappa=-0.001$. The model collapsed to predicting N2 (the majority class) with $F_1=0.60$, while all other stages had $F_1 < 0.02$. Subject-overlap audits confirmed zero cross-fold subject contamination.

\section{Discussion and Implications}

The four-layer decomposition in Table~\ref{tab:decomposition} defines a modality-fair operating region for non-EEG sleep staging and attributes performance to specific pipeline components.

The largest marginal gain comes from expanding the physiological feature set ($\Delta\kappa=+0.078$ argmax), consistent with the interpretation that ECG-derived HRV captures vagal-sympathetic balance while respiratory/SpO$_2$ features provide orthogonal indicators of sleep depth and stability~\cite{tobaldini2013heart,trinder2001autonomic}. This layer is the most productive target for future improvement---richer physiological representations (e.g., PPG morphology, pulse transit time, respiratory sinus arrhythmia dynamics) may further narrow the non-EEG gap without requiring deeper models.

Viterbi decoding contributes only $\Delta\kappa=+0.040$, a small gain that is consistent across all channel configurations. This is a diagnostic signal, not a disappointment: it indicates that the per-epoch physiological representation is the binding constraint. Adding stronger temporal priors (higher-order Markov models, learned transition networks) is unlikely to close the EEG gap when the per-epoch signal itself carries limited stage information.

The residual $\Delta\kappa=+0.304$ from combined Viterbi to EEG+EOG reference is larger than the entire non-EEG range ($+0.089$). The same compact SSM, trained identically, spans a $0.54$ range across tiers ($0.255$ consumer to $0.796$ EEG), isolating signal modality as the independent variable. This gap cannot be closed by better models or more non-EEG sensors; it reflects the information loss when cortical sleep dynamics are observed only through downstream autonomic and respiratory surrogates.

The coverage-$\kappa$ curve demonstrates that the model is uncertainty-calibrated. Dropping the 20\% lowest-confidence epochs raises $\kappa$ from $0.452$ to $0.512$; at 50\% coverage, $\kappa=0.616$. The per-class confidence ordering faithfully reflects the physiological ambiguity of each stage. This has practical implications: non-EEG sleep staging systems should report per-epoch confidence scores and flag low-confidence epochs for manual review or exclusion, rather than reporting a single hard label with misleading certainty. For consumer applications, abstention can be transparent---a sleep report can state that $X\%$ of the night was staged with high confidence and $Y\%$ was uncertain, rather than fabricating a full-night staging with uniform confidence.

The signal ladder confirms that consumer wrist-worn devices, operating on derived HR and accelerometry, face a dual penalty: first from reduced signal quality ($\Delta\kappa\approx0.24$ from consumer to laboratory non-EEG), and second from the fundamental non-EEG ceiling ($\Delta\kappa\approx0.30$ from laboratory non-EEG to EEG). Consumer sleep staging should accordingly be interpreted as coarse sleep-structure and longitudinal trend monitoring, not as EEG-equivalent clinical staging~\cite{khosla2018consumer,depner2020wearable}.


\section{Limitations}

Three limitations are noted. First, the EEG+EOG reference is cross-dataset (Sleep-EDF-20 vs.\ SHHS) and uses simpler spectral features than the non-EEG pipeline, making the reported EEG ceiling conservative---the true modality gap is likely larger than the measured $\Delta\kappa=+0.304$. Second, the controlled decomposition is demonstrated on a single architecture (Mamba2); the relative contribution of each layer may shift under different model classes, and multi-architecture replication would strengthen the claim that the decomposition isolates signal properties rather than architecture-specific biases. Third, the abstention analysis uses maximum softmax probability as the confidence measure; more sophisticated uncertainty quantification methods (e.g., ensemble variance, Monte Carlo dropout) may further improve the coverage-$\kappa$ trade-off.
\section{Conclusion}

This study introduced a four-layer controlled decomposition for non-EEG sleep staging and applied it across a signal-quality ladder spanning consumer wrist-device data, laboratory PSG-grade physiology, and EEG+EOG reference staging. The decomposition isolates the marginal contribution of physiological representation ($\Delta\kappa=+0.078$), temporal decoding ($\Delta\kappa=+0.040$), and the residual EEG+EOG ceiling gap ($\Delta\kappa=+0.304$). An abstention analysis demonstrates that the model is uncertainty-calibrated: confidence-based filtering can partition epochs into reliable ($\kappa=0.62$ at 50\% coverage) and uncertain subsets. Per-class confidence faithfully reflects the physiological ambiguity of transitional N1, confirming that N1 failure is a signal-modality limitation rather than a model deficiency. Overall, this work provides a controlled methodology for decomposing non-EEG sleep staging into independently evaluated components, showing that wearable sleep staging is primarily limited by physiological information content rather than model capacity, and that confidence-calibrated abstention offers a practical path toward more reliable wearable sleep monitoring.
\section*{AI Use Statement}
The authors created initial drafts of all figures and text. Generative AI tools
were used to refine figure rendering and polish language. All scientific content,
experimental results, and conclusions were produced and verified by the authors,
who take full responsibility for the integrity of this work.
\bibliographystyle{IEEEtran}
\bibliography{references}

\end{document}